\documentclass[letterpaper]{article} %
\usepackage{aaai2027}  %
\usepackage[hyphens]{url}  %
\usepackage{graphicx} %
\usepackage{natbib}  %
\usepackage{caption} %
\usepackage{algorithm}
\usepackage{algorithmic}

\usepackage{newfloat}
\usepackage{listings}
\DeclareCaptionStyle{ruled}{labelfont=normalfont,labelsep=colon,strut=off} %
\floatstyle{ruled}
\newfloat{listing}{tb}{lst}{}
\floatname{listing}{Listing}

\usepackage{booktabs}

\nocopyright %

\title{Reconfiguration of Temporal Networks under Reachability Constraints}
\author{
    Othon Michail\textsuperscript{\rm 1},
    George Skretas\textsuperscript{\rm 2},
    Georg Tennigkeit\textsuperscript{\rm 2}, %
    Shaily Verma\textsuperscript{\rm 3}
}
\affiliations{
    \textsuperscript{\rm 1}University of Patras, Patras, Greece\\
    \textsuperscript{\rm 2}Hasso Plattner Institute, University of Potsdam, Potsdam, Germany\\
    \textsuperscript{\rm 3}Indian Institute of Technology, Jodhpur, India

    othon.michail@upatras.gr, georgios.skretas@hpi.de, georg.tennigkeit@hpi.de, shailyverma048@gmail.com
}

\usepackage{amsthm}
\usepackage{graphicx} 
\usepackage{xspace}
\usepackage{amssymb}
\usepackage{amsthm}
\usepackage{amsmath,mathtools}
\usepackage{cleveref}

\newtheorem{theorem}{Theorem}
\newtheorem{lemma}{Lemma}

\newcommand{\bigparagraph}[1]{\vspace{0.4em}\noindent\textbf{#1}}

\newcommand{\gcal}{\ensuremath{\mathcal{G}}\xspace}

\newcommand{\PSPACE}{{\ensuremath{\mathtt{PSPACE}}\xspace}}

\newcommand{\natN}{\ensuremath{\mathbb{N}}\xspace}

\newcommand{\tuple}[1]{\ensuremath{\langle {#1} \rangle}\xspace}

\newcommand{\reconf}[1]{\ensuremath{\xleftrightarrow{#1}}\xspace}

\newcommand{\sourcereach}[1]{\ensuremath{\ensuremath{#1}\mathtt{SR}}\xspace}
\newcommand{\disjointpathrec}{\textsc{DPR}\xspace}
\newcommand{\directededgedisjointpathrec}{\textsc{DEDPR}\xspace}

\begin{document}

\maketitle

\begin{abstract}
Temporal networks model dynamic systems in which edges represent interactions and labels specify when these interactions occur. Examples include transportation networks, time-sensitive communication networks, and industrial control systems. In many such applications, an existing temporal network must be transformed into a desired one through a sequence of atomic modifications while maintaining essential functionality throughout the transformation. We formalize the time label reconfiguration problem and provide a theoretical framework for reasoning about such transformation processes.

As temporal reachability is a central functionality in many temporal networks, we study a reconfiguration problem on directed temporal graphs subject to temporal reachability constraints.
We are given a static graph with a designated set of sources, along with two labeling functions indicating an availability time for every edge. The goal is to transform one labeling into the other by changing the label of a single edge at a time while maintaining temporal reachability of the sources throughout.
Our results reveal a sharp complexity transition: the problem is polynomial-time solvable for a single source but becomes PSPACE-hard with two sources. We also show that if the static graph is acyclic or an \emph{almost-tournament graph}, then all valid labelings can be reconfigured into each other. Our proofs are constructive and yield polynomial-time algorithms.

\end{abstract}

\section{Introduction}

Many dynamic systems are naturally modeled as temporal networks, where edges represent interactions, and their labels specify the times at which they occur. Examples include transportation networks, communication networks, and industrial control systems. In practice, such schedules evolve due to disruptions, changing demands, or new operational requirements. Since these systems typically remain operational during the update process, one schedule cannot simply be replaced by another. Instead, the schedule must be transformed incrementally while maintaining the required operational guarantees throughout the transformation.

This motivates the following general question: given two feasible temporal labelings of the same graph, can one be transformed into the other by changing the label of a single edge at a time while maintaining feasibility throughout the entire transformation? We refer to this as the time label reconfiguration problem.

Such problems are present in airline disruption recovery, where flight schedules must be adapted in response to disruptions while the transportation network continues to operate. Rather than replacing the schedule at once, airlines perform incremental updates that maintain operational feasibility throughout the transformation \cite{Lettovsky2000,Clausen2010}. Viewing flights as temporal edges and departure times as edge labels, this naturally corresponds to modifying time labels while preserving the functionality of the temporal network.

A similar challenge arises in Time-Sensitive Networking (TSN), where deterministic communication is achieved by assigning transmission times to communication streams over a fixed network topology \cite{Zhang2024}. As communication demands evolve or devices are added and removed, the schedule must be adapted without disrupting ongoing communication. Consequently, TSN research has studied incremental schedule migration, where a running schedule is transformed into a new one while preserving deterministic communication guarantees during the transformation \cite{Gaertner2022,Gaertner2023}. Our model provides a graph-theoretic abstraction of this process by considering a fixed communication topology and modifying one transmission at a time.

\subsection{Our Contribution}

In this paper, we study time label reconfiguration under temporal reachability constraints, as a common criterion for network robustness is that designated vertices must remain able to reach the entire network. Given two temporal labelings of the same directed graph that both allow a designated set of source vertices to reach every remaining vertex, we ask whether one labeling can be transformed into the other by repeatedly changing the label of a single edge while preserving this reachability property throughout the transformation. Unlike traditional temporal graph problems, reconfiguration requires reasoning simultaneously about the initial labeling, the target labeling, and every intermediate labeling, yielding a large implicit state space whose transformations are constrained by temporal reachability.

Our main results establish a sharp complexity transition as a function of the number of sources. For a single source, we show that the reconfiguration problem can be decided in polynomial time, and that a shortest reconfiguration sequence can be computed efficiently. In contrast, allowing two sources already makes the decision problem PSPACE-hard. Since reconfigurability is a desirable property, we investigate which graph properties guarantee reconfigurability. We observe that cyclic dependencies between source-to-vertex paths prevent reconfiguration. This naturally motivates investigating graph classes in which such dependencies are absent or highly restricted.
We first consider directed acyclic graphs, where cycles are eliminated entirely, and show that every instance is reconfigurable by providing a polynomial-time algorithm that constructs a reconfiguration sequence. We then investigate whether this result extends to the broader class of chordal graphs, whose cycles are highly constrained. Surprisingly, we prove that a directed analogue of chordality alone is insufficient by exhibiting an infinite family of chordal graphs that are not reconfigurable. Finally, we identify a positive graph class by considering \emph{almost-tournament} graphs, obtained by removing the edge between the two source vertices of a tournament. The dense connectivity of these graphs provides sufficient routing flexibility to resolve cyclic dependencies, allowing us to prove that every instance is reconfigurable.

\subsection{Related Work}

\bigparagraph{Reconfiguration on Temporal Structures.}
Some reconfiguration problems have been considered on temporal graphs before. In \emph{Temporal Arborescence Reconfiguration}~\cite{ito2023reconfiguration,dondi2024complexity}, a temporal arborescence is a rooted directed spanning tree where consecutive edges have increasing time labels. This reconfiguration problem is always solvable if the given arborescences have the same root, and still efficiently decidable otherwise. Unlike in our problem, the temporal graph is always the same, and instead a subset of edges is maintained that ensures reachability.
\citet{DBLP:conf/algosensors/DavotEL25} introduce \emph{temporally satisfying reconfiguration} for generic vertex selection problems: Here, we need to find a separate solution for every time step such that every solution can be turned into the solution for the next step via parallel token shifting. In other words, reconfiguration steps coincide with time steps, which is a significant difference from the common reconfiguration framework~\cite{DBLP:journals/algorithms/Nishimura18}.

\bigparagraph{Reconfiguration under Static Connectivity Constraints.}
Substantial prior work has studied reconfiguring paths (specifically shortest paths) in static graphs \cite{DBLP:conf/aaai/GajjarJ0L22,Bonsma_2017,DBLP:journals/tcs/KaminskiMM11}.
\citet{DBLP:conf/isaac/HanakaIKOS24} investigate reconfiguration of a sequence of disjoint spanning trees and provide an algorithm for the generalized problem of matroid basis sequence reconfiguration. 
Closer to our chosen constraint is the reconfiguration of rooted arborescences~\cite{DBLP:journals/tcs/ItoIKNOW23};
Other reconfiguration problems concerning connectivity focus on connected partitions~\cite{DBLP:journals/tcs/AkitayaKKST22}, cliques~\cite{DBLP:journals/dam/ItoOO23}, and routing tables in software-defined networks~\cite{DBLP:conf/hotnets/ReitblattFRW11}.

\bigparagraph{Time Label Modification.}
Our work is motivated by network optimization problems that determine a labeling for a given temporal graph that improve an existing labeling regarding some metric. Commonly, labels are shifted to either increase connectivity~\cite{DBLP:conf/ijcai/DeligkasES23,DBLP:journals/tcs/EnrightLMP26} or to decrease it~\cite{DBLP:journals/jcss/MolterRZ24,DBLP:journals/iandc/DeligkasP22}.
Optimization of labelings traces its roots to \citet{DBLP:journals/jcss/KempeKK02}, who studied temporal label inference, and \citet{DBLP:journals/algorithmica/MertziosMS19}, who investigated the design of multi-labeled temporal graphs minimizing labeling complexity under reachability constraints.

\section{Preliminaries}
    For $n\in \natN$, we denote the set ${\{1,\dots,n\}}$ by $[n]$.
    A directed \textit{temporal graph} $\gcal=(V,E,\lambda)$ consists of a directed static graph $G=(V,E)$, called the \textit{footprint}, and a labeling function $\lambda: E\rightarrow \natN_0$. For each edge $e\in E$, $\lambda(e)$ is the time step at which $e$ is active. Unless specified otherwise, all labelings are \emph{simple}, %
    meaning every edge has exactly one label.
    Unless specified otherwise, all graphs are directed.
    
    A \textit{temporal path} is a sequence of edges $\tuple{e_i}$ that forms a path where the time labels $\tuple{\lambda(e_i)}$ are strictly increasing. %
    If there exists a temporal path from $u$ to $v$, we say \emph{$u$ reaches $v$}. %
    We denote the set of vertices that $u$ reaches in \gcal as $R^\gcal(u)$.
    
    \bigparagraph{$k$-Source Reachability.}
    Given a temporal graph $\gcal=(V\cup S, E, \lambda)$ with a designated set of source vertices $S$ where $|S|=k$, we say that \gcal satisfies \emph{$k$-source reachability} (\sourcereach{k}) if every source can reach all non-source vertices, that is, if $V\subseteq R^\gcal(s)$ for all $s\in S$. For simplicity, we may also say that a labeling $\lambda^*$ for a static graph satisfies \sourcereach{k} if the temporal graph $(V\cup S, E, \lambda^*)$ satisfies \sourcereach{k}.

    \bigparagraph{Time Label Reconfiguration.}
    A reconfiguration step consists of changing the label of a single edge. Formally, we say that a labeling $\lambda$ can be turned into a labeling $\lambda'$ in one step iff there is an $e\in E$ such that $\forall e'\neq e\colon \lambda(e')=\lambda'(e')$.
    
    Based on this, we define reconfigurability with respect to (wrt) some property $\Pi$: %
    Let $G=(V, E)$ be a static graph, and let $\lambda_1, \lambda_2\colon E\rightarrow \natN$ be two labelings for $G$ such that both $\lambda_1$ and $\lambda_2$ satisfy $\Pi$ on $G$. We say that $\lambda_1$ is reconfigurable into $\lambda_2$ with respect to $\Pi$ (or $\lambda_1\reconf{\Pi}\lambda_2$) iff there is a sequence of labelings $\tuple{\lambda_1=\lambda^1, \lambda^2, \dots \lambda^\ell=\lambda_2}$ such that $(V, E, \lambda^i)$ satisfies $\Pi$ and $\lambda^i$ can be turned into $\lambda^{i+1}$ in one step for every $i\in [\ell-1]$. Note that reconfiguration sequences are reversible, hence reconfigurability is a symmetric relation.

\section{Single-Source Reachability}

This problem is related to the problem of time-respecting arborescence reconfiguration with a fixed source~\cite{ito2023reconfiguration}. That problem is always solvable, and reconfiguration wrt \sourcereach{1} is possible in a similar manner. In fact, we can efficiently find a reconfiguration sequence of minimal length.

\begin{theorem}\label{thm:1source_reach}
    Given a static graph $G=(V\cup \{s\}, E)$ with a single source $s$, we have $\lambda_1\reconf{\sourcereach{1}}\lambda_2$, for any labelings $\lambda_1, \lambda_2$ satisfying \sourcereach{1}, and a shortest reconfiguration sequence can be computed in polynomial time. %
\end{theorem}
\begin{proof}

We first determine a canonical labeling $\lambda_c$, then we describe how to reconfigure both $\lambda_1$ and $\lambda_2$ into $\lambda_c$.

Construct the merged temporal graph $\gcal_{1+2}=(V\cup \{s\}, E, \lambda_1\cup \lambda_2)$ where $\lambda_1\cup \lambda_2$ returns the set $\{\lambda_1(e), \lambda_2(e)\}$ for every $e\in E$ (this temporal graph has up to two labels per edge). Then run a temporal Dijkstra~\cite{DBLP:journals/networks/HalpernP74,DBLP:journals/ijfcs/XuanFJ03} calculating foremost journeys on $\gcal_{1+2}$ starting from $s$ (this variant of Dijkstra traverses outgoing edges from earliest to latest, ignoring those occurring before the arrival time at a vertex). Record the order in which edges are traversed, and for every edge $e$ traversed with label $t$, define $\lambda_c(e):=t$. At the end, for every edge $e$ where $\lambda_c(e)$ is not yet defined, set $\lambda_c(e)=\lambda_1(e)$.

\begin{figure}[h]
    \centering
    \includegraphics[width=\linewidth]{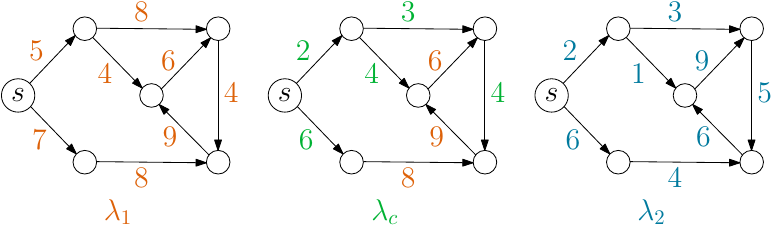}
    \caption{Given two labelings $\lambda_1, \lambda_2$, we determine a canonical labeling $\lambda_c$ based on the labels used by a temporal Dijkstra on the merged graph.}
    \label{fig:1SR_example}
\end{figure}

To reconfigure $\lambda_1$ into $\lambda_c$, we iterate through the edges in the previously recorded Dijkstra order and for each edge $e$ change its label to $\lambda_c(e)$. We show that \sourcereach{1} is satisfied after every such operation: If this operation reduces the label of $e$, then for any previous temporal path $P$ from $s$ to some $v$ passing through $e$, we now have a temporal path from $s$ to $v$ by following the temporal Dijkstra up to $e$ and then appending the suffix of $P$ after $e$. If the operation increases the label of $e$, then we know that $e$ was not traversable from $s$ before, since otherwise its current smaller label would have been used in the temporal Dijkstra. Hence changing its label cannot reduce reachability.

To reconfigure $\lambda_2$ into $\lambda_c$, we proceed likewise, but then we also change all other labels (those not used in the Dijkstra) to their value in $\lambda_c$, in any order. These operations cannot reduce reachability as the labels of the Dijkstra are sufficient on their own to satisfy \sourcereach{1}.

To show $\lambda_1\reconf{\sourcereach{1}}\lambda_2$, reverse the reconfiguration sequence from $\lambda_2$ to $\lambda_c$ and append it to the reconfiguration sequence from $\lambda_1$ to $\lambda_c$. Note that every reconfiguration step resolves a difference between $\lambda_1$ and $\lambda_2$, and at least one step for every such difference is required. Hence this sequence has the minimum length.%
\end{proof}

\section{Hardness of Two-Source Reachability}

We now continue to the setting of $k=2$, where there are two sources that both need to keep reaching all other vertices throughout reconfiguration. Immediately, we find that not all instances of this problem are reconfigurable.

\begin{lemma}
    Reconfiguration wrt \sourcereach{2} is not always possible.
\end{lemma}
\begin{proof}

We give a counterexample in \Cref{fig:2SR_no_instance}. Note that $\lambda_1$ and $\lambda_2$ differ only in two edges. This instance can be generalized to any number of vertices, for example, by subdividing the edge $(c, s_1)$ into a path and increasing labels accordingly.

\begin{figure}[h]
    \centering
    \includegraphics[width=\linewidth]{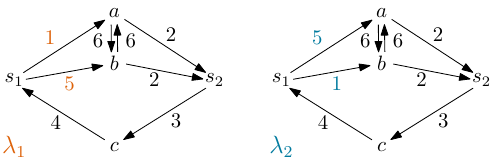}
    \caption{An instance where reconfiguration is not possible. Labels where $\lambda_1$ and $\lambda_2$ differ are highlighted.}
    \label{fig:2SR_no_instance}
\end{figure}

The reason this instance is not reconfigurable is because of the relation between three specific edges: At least one of $(s_1, a)$ and $(s_1, b)$ must have a smaller label than $(s_2, c)$, or else $s_1$ could not reach $c$. Simultaneously, at least one of $(s_1, a)$ and $(s_1, b)$ must have a larger label than $(s_2, c)$, or else $s_2$ could not reach $a$ or $b$. Both of these inequalities must be met in every step of a reconfiguration sequence wrt \sourcereach{2}.

In Figure \ref{fig:2SR_no_instance}, we observe that $\lambda_1((s_1, a)) < \lambda_1((s_2, c)) < \lambda_1((s_1, b))$, and that $\lambda_2((s_1, b)) < \lambda_2((s_2, c)) < \lambda_2((s_1, a))$, both of which satisfy the above inequalities. Indeed, these are the only two orderings of the critical edges for which there is a labeling satisfying \sourcereach{2}, as all other orderings would violate one of the previous inequalities. Because a single operation cannot completely flip the ordering in $\lambda_1$ to that in $\lambda_2$, and intermediate steps would not satisfy \sourcereach{2}, reconfiguration is not possible.
\end{proof}

\subsection{PSPACE-hardness}

The above instance prevents reconfiguration via two edges where one needs to occur earlier than some \emph{pivot edge}, and the other needs to occur later than the same pivot edge, and their roles cannot be swapped by any one operation. If there were no direct edges from $s_1$ to $a$ and $b$, then we would instead require one path with small labels, and another with large labels, with their edge sets being disjoint as a result. The reconfiguration variant of this is closely related to \textsc{Disjoint $s$-$t$ Paths Reconfiguration} or \disjointpathrec~\cite{DBLP:journals/talg/ItoIK0MNOO25}. Indeed, the \PSPACE-hardness of \disjointpathrec extends to our problem.

We recount the definition of \disjointpathrec: We are given an undirected graph $G$ with source vertex $s$ and target vertex $t$, as well as two \emph{linkages}, that is, two pairs of internally vertex-disjoint paths $(P_1, P_2)$ and $(Q_1, Q_2)$ from $s$ to $t$ (the authors parameterize the number of paths per linkage with $k$, but we only need $k=2$). As an operation, a pair of paths $(P_1, P_2)$ can be turned into $(P', P_2)$ or $(P_1, P')$ where $P'$ is any $s$-$t$ path. The pair of paths needs to be internally vertex-disjoint throughout. They showed \PSPACE-hardness for deciding whether $(P_1, P_2)$ can be reconfigured into $(Q_1, Q_2)$ in this way.

As we use directed graphs, and time labels intuitively partition edges rather than vertices, we define \textsc{Directed Edge-Disjoint $s$-$t$ Paths Reconfiguration} (or \directededgedisjointpathrec) as a variant of \disjointpathrec. We first show \PSPACE-hardness of this variant. %

\begin{lemma}
    \directededgedisjointpathrec is \PSPACE-hard.
\end{lemma}
\begin{proof}
    We reduce from \disjointpathrec. Given a \disjointpathrec instance $(G=(V, E), s, t, (P_1, P_2), (Q_1, Q_2))$, create a \directededgedisjointpathrec instance $(G'=(V', E'), s, t, (P_1', P_2'), (Q_1', Q_2'))$ as follows:
    \begin{enumerate}
        \item For each $v\in V\setminus\{s, t\}$, create vertices $v^{in}, v^{out}$ along with a directed edge $(v^{in}, v^{out})$. Also create vertices $s, t$.
        \item For each edge $\{u, v\}\in E$ that is not incident to $s$ or $t$, create two directed edges $(u^{out}, v^{in}), (v^{out}, u^{in})$.%
        \item For each $\{s, v\}\in E$, create the directed edge $(s, v^{in})$. For each $\{v, t\}\in E$, create the directed edge $(v^{out}, t)$.
        \item Create $P_1'$ by copying $P_1$ and replacing each $v \in V\setminus\{s, t\}$ with the sequence $v^{in}, v^{out}$. %
        Do likewise for $P_2', Q_1', Q_2'$.
    \end{enumerate}

    \begin{figure}[h]
        \centering
        \includegraphics[width=0.75\linewidth]{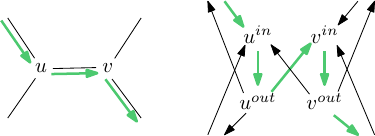}
        \caption{For reducing \disjointpathrec to its directed and edge-disjoint variant, we split every vertex into an in- and out-vertex. Original paths are rerouted through both of these vertices.}
        \label{fig:disjoint_paths_reconf_to_directed_edgedisj}
    \end{figure}

    Refer to \Cref{fig:disjoint_paths_reconf_to_directed_edgedisj} for an illustration. We prove a bijection $b$ from the $s$-$t$ paths in $G$ to those in $G'$, where furthermore a pair of paths $(P_1, P_2)$ in $G$ is internally vertex-disjoint if and only if the pair of paths $(b(P_1), b(P_2))$ is edge-disjoint. This implies that reconfiguration steps are equivalent in both instances, proving the reduction.

    We define $b$ as per step 4 of the above construction. $b$ is injective by definition. To prove surjectivity, we observe that any $v^{in}$ only has one outgoing edge to $v^{out}$, and any $v^{out}$ has only one incoming edge from $v^{in}$. Hence any $s$-$t$ path $P'$ in $G'$ must be made up of a sequence of $v^{in}, v^{out}$ pairs. So $b^{-1}(P')$ is the same path in $G$ where every consecutive $v^{in}, v^{out}$ is replaced with $v$.

    Let $(P_1, P_2)$ be internally vertex-disjoint in $G$. Then $(b(P_1), b(P_2))$ are internally vertex-disjoint in $G'$ by definition of $b$ and thus also edge-disjoint. Inversely, let $(P_1, P_2)$ not be internally vertex-disjoint, so they share a vertex $v$. Then $(b(P_1), b(P_2))$ share the edge $(v^{in}, v^{out})$ and are therefore not edge-disjoint. This means that a reconfiguration step replacing $(P_1, P_2)$ in $G$ wlog with $(P', P_2)$ corresponds exactly to a reconfiguration step replacing $(b(P_1), b(P_2))$ in $G'$ with $(b(P'), b(P_2))$. 
\end{proof}

We now use this variant of \disjointpathrec to prove hardness of time label reconfiguration wrt \sourcereach{2}. The main idea is to insert an instance of \directededgedisjointpathrec into the construction in \Cref{fig:2SR_no_instance} to replace the edges labeled $1$ and $5$.

\begin{theorem}\label{thm:2sr_pspace_hard}
    Given a static graph $G=(V\cup \{s_1, s_2\}, E)$ with two sources and two labelings $\lambda_1, \lambda_2$ satisfying \sourcereach{2}, it is \PSPACE-hard to decide whether $\lambda_1\reconf{\sourcereach{2}}\lambda_2$.
\end{theorem}
\begin{proof}
We reduce from \directededgedisjointpathrec. Given an instance $(G_P=(V_P,E_P), s_1, t, (P_1, P_2), (Q_1, Q_2))$, we construct $\gcal$ and $\lambda_1, \lambda_2$ as follows (see also \Cref{fig:2SR_pspace_hard_construction}):
\begin{enumerate}
    \item Duplicate $G_P$ and add vertices $s_2$ and $h$ and edges $(t, s_2)$, $(s_2, h)$, and $(h, s_1)$. Define the set of sources as $\{s_1, s_2\}$.
    \item For every $v\in V_P\setminus\{s_1, t\}$, add a \emph{back-edge} $(t, v)$. Note that $s_1$-$t$ paths cannot contain back-edges.
    \item Let $|V_P|=n$. For any pair of edge-disjoint $s_1$-$t$ paths $(X_1, X_2)$, let the labeling $\lambda_{X_1, X_2}$ be defined as follows:
    \begin{enumerate}
        \item For every $e_1\in X_1$, if $e_1$ is the $i$-th edge along $X_1$, define $\lambda_{X_1, X_2}(e_1)=i$.
        \item Define $\lambda_{X_1, X_2}((t, s_2))=2n$, $\lambda_{X_1, X_2}((s_2, h))=2n+1$, and $\lambda_{X_1, X_2}((h, s_1))=2n+2$.
        \item For every $e_2\in X_2$, if $e_2$ is the $i$-th edge along $X_2$, define $\lambda_{X_1, X_2}(e_2)=3n+i$.
        \item For every back-edge $e_b$, define $\lambda_{X_1, X_2}(e_b)=5n$.
        \item For all other edges $e'$, define $\lambda_{X_1, X_2}(e')=0$.
    \end{enumerate}
    \item Define $\lambda_1=\lambda_{P_1, P_2}$ and $\lambda_2=\lambda_{Q_1, Q_2}$.
\end{enumerate}

\begin{figure}[h]
    \centering
    \includegraphics[width=\linewidth]{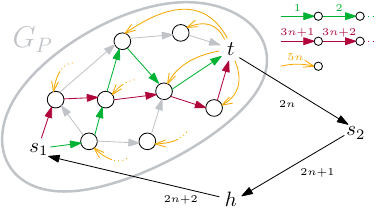}
    \caption{In the construction used for \Cref{thm:2sr_pspace_hard}, an instance of \directededgedisjointpathrec is embedded into a 4-cycle. Two edge-disjoint $s_1$-$t$ paths are labeled incrementally starting from $1$ and $3n+1$ respectively. Back-edges go out from $t$ and have label $5n$. All other edges (gray) have label $0$.}
    \label{fig:2SR_pspace_hard_construction}
\end{figure}

This instance can be constructed in linear time. We first show that for any pair of edge-disjoint paths $(X_1, X_2)$, $\lambda_{X_1, X_2}$ satisfies \sourcereach{2} on $\gcal$. $s_1$ can reach $t$ via $X_1$, and from $t$ it can reach all other vertices in $V_P$ via back-edges as well as $s_2$ and $h$. $s_2$ can reach $s_1$ via $h$ and from there follow $X_2$ to $t$ and continue along back-edges to all other vertices in $V_P$. In particular, $\lambda_1$ and $\lambda_2$ satisfy \sourcereach{2} on $\gcal$.

\bigparagraph{\directededgedisjointpathrec $\Rightarrow$ $\lambda_1\reconf{\sourcereach{2}}\lambda_2$}

\noindent Let $\tuple{(P_1=P_1^1, P_2=P_2^1), (P_1^2, P_2^2), \dots (P_1^\ell=Q_1, P_2^\ell=Q_2)}$ be a reconfiguration sequence from $(P_1, P_2)$ to $(Q_1, Q_2)$. We describe for any $i\in [\ell-1]$ how to reconfigure $\lambda_{P_1^i, P_2^i}$ into $\lambda_{P_1^{i+1}, P_2^{i+1}}$. This implies by transitivity that $\lambda_1=\lambda_{P_1, P_2}$ can be reconfigured into $\lambda_2=\lambda_{Q_1, Q_2}$.

If $P_1^i=P_1^{i+1}$ and $P_2^i=P_2^{i+1}$, then %
their implied labelings are also identical and we are done. Otherwise assume wlog that $P_1^i\neq P_1^{i+1}$. We reconfigure as follows:
\begin{enumerate}
    \item Increase the labels along all edges of $P_1^i$ by $n$, from the end to the beginning of the path. This never violates \sourcereach{2} as the path remains temporal and switching onto other edges remains possible.
    \item Iterate along the edges of $P_1^{i+1}$ from beginning to end: Change the label of its $j$-th edge to $j$. None of these changes break \sourcereach{2}: If an edge had label $0$ before, it was not needed for reachability as described earlier. If it had another label, it must be on $P_1^i$ ($P_1^{i+1}$ cannot contain back-edges and is edge-disjoint from $P_2^i$ by definition). Then this operation decreases the label of the edge, but it is still traversable by a path starting from $s_1$, so the reachability of $s_1$ does not decrease.
    \item In any order, set the labels of edges in $P_1^i \setminus P_1^{i+1}$ to $0$. Because $P_1^{i+1}$ and $P_2^i=P_2^{i+1}$ are sufficient to ensure \sourcereach{2}, none of these operations break reachability. The resulting labeling is exactly $\lambda_{P_1^{i+1}, P_2^{i+1}}$.
\end{enumerate}

\bigparagraph{$\lambda_1\reconf{\sourcereach{2}}\lambda_2$ $\Rightarrow$ \directededgedisjointpathrec}

\noindent Given a reconfiguration sequence of labelings $\tuple{\lambda_1=\lambda^1, \lambda^2, \dots \lambda^\ell=\lambda_2}$, let us first define the following sets: For $i\in [\ell]$, let $E_{early}^i=\{e\in E_P \mid \lambda^i(e) < \lambda^i((s_2, h))\}$ and $E_{late}^i=\{e\in E_P \mid \lambda^i(e) > \lambda^i((s_2, h))\}$. These sets are disjoint by definition.

We observe how $E_{early}^{i+1}$ and $E_{late}^{i+1}$ change compared to $E_{early}^i$ and $E_{late}^i$ based on the relabeling operation applied between them:
If the label of an edge in $E_P$ is increased, or if the label of $(s_2, h)$ is decreased, we have $E_{late}^i\subseteq E_{late}^{i+1}$.
If the label of an edge in $E_P$ is decreased, or if the label of $(s_2, h)$ is increased, we have $E_{early}^i\subseteq E_{early}^{i+1}$.
In any case, at least one of $E_{late}^i\subseteq E_{late}^{i+1}$ and  $E_{early}^i\subseteq E_{early}^{i+1}$ holds.

We will construct a reconfiguration sequence $\tuple{(P_1=P_1^1, P_2=P_2^1), \dots (P_1^{\ell+2}=Q_1, P_2^{\ell+2}=Q_2)}$ with the following invariant:  $P_1^i\subseteq E_{early}^i$, and $P_2^i \subseteq E_{late}^i$. This invariant holds for $i=1$ by construction.

Now for any $i\in \{2, \dots \ell\}$, we observe that there must be a temporal $s_1$-$t$ path $P_1^i$ in $(\gcal, \lambda^i)$ that occurs entirely before $\lambda^i((s_2, h))$, or else $s_1$ could not reach $h$. Likewise, there must be a temporal $s_1$-$t$ path $P_2^i$ in $(\gcal, \lambda^i)$ that occurs entirely after $\lambda^i((s_2, h))$, or else $s_2$ could not reach $t$.
Hence $P_1^i$ only contains edges of $E_{early}^i$ and $P_2^i$ only contains edges of $E_{late}^i$, and they are edge-disjoint by definition. Because we have either $E_{late}^{i-1}\subseteq E_{late}^{i}$ or $E_{early}^{i-1}\subseteq E_{early}^{i}$, we can choose $P_1^i=P_1^{i-1}$ or $P_2^i=P_2^{i-1}$. This guarantees that $(P_1^{i-1},P_2^{i-1})$ can be reconfigured into $(P_1^i,P_2^i)$ in one step.

For $(P_1^\ell, P_2^\ell)$, we have edge-disjoint paths where $P_1^\ell\subseteq E_{early}^\ell$, and $P_2^\ell \subseteq E_{late}^\ell$. These might not be the same as $(Q_1, Q_2)$, but both follow the invariant for $i=\ell$. To complete the reconfiguration sequence, we define $(P_1^{\ell+1}, P_2^{\ell+1})=(Q_1, P_2^{\ell})$ and $(P_1^{\ell+2}, P_2^{\ell+2})=(Q_1, Q_2)$.
\end{proof}

\section{Reconfigurability by Footprint Properties}

To better understand the structural properties that prevent reconfiguration, we investigate graph properties that guarantee reconfiguration if the footprint exhibits them.

\subsection{DAGs}

The no-instance in \Cref{fig:2SR_no_instance} implies a dependency between two labels, which relies on these edges being on the same cycle. We show that reconfiguration is always possible for any number of sources if the footprint is acyclic, and a reconfiguration sequence can be computed efficiently. To do so, we give a canonical labeling that any other labeling can be reconfigured to.

\begin{theorem}\label{thm:2source_reach_atl_directed_on_dags}
        Given an acyclic graph $G=(V\cup S, E)$ with for $|S|=k$, we have $\lambda_1\reconf{\sourcereach{k}}\lambda_2$ for any labelings $\lambda_1, \lambda_2$ satisfying \sourcereach{k}, and a reconfiguration sequence can be computed in linear time.
\end{theorem}
\begin{proof}
    Find a topological order of the DAG; let $i(v)$ be the index of vertex $v$ in the order. We determine a canonical labeling $\lambda_c$: For every edge $e=(u, v)$, set $\lambda_c(e)=i(u)$:

    We show reconfigurability of any labeling into $\lambda_c$, which in turn shows $\lambda_1\reconf{\sourcereach{k}}\lambda_c\reconf{\sourcereach{k}}\lambda_2$. To reconfigure any labeling $\lambda$ into $\lambda_c$, first shift up all labels in $\lambda$ by $n=|V\cup S|$. This can be done by repeatedly adding $n$ to the highest label that has not yet been increased. Now iterate through the vertices in topological order: For each $v$ in the order, set the label of every outgoing edge equal to its label in $\lambda_c$. Because we previously shifted up the labels in $\lambda$ by $n$ and $\lambda_c$ only contains labels less than $n$, we are only decreasing labels.
    
    Any existing temporal path $P$ starting from a source remains temporal after each of these operations: Assume $P$ is not temporal after decreasing the label of an edge $(v, w)$, then the edge $(u, v)$ preceding $(v, w)$ on $P$ must have a higher or equal label. But because $(u, v)$ exists, $u$ must be before $v$ in the topological order, so we already decreased the outgoing edges of $u$ to $i(u)$, which is smaller than $i(v)$ and contradicts our assumption.

    Finding a topological ordering of $G$, shifting up all labels, and assigning the labels of $\lambda_c$ each take linear time.
\end{proof}

\subsection{Chordal Graphs}

The no-instance in \Cref{fig:2SR_no_instance} relies on induced 4-cycles, and one can show that if either 4-cycle had a chord, the instance would be reconfigurable due to the resulting alternative path to a vertex. This raises the question whether chordal graphs can guarantee reconfiguration. We consider \emph{disoriented-chordal graphs} as an adaptation to directed graphs: A directed graph is disoriented-chordal if the disoriented graph (i.e.\ the same graph with undirected edges) is chordal. \footnote{Another plausible adaptation would be the class of graphs that contain no \emph{directed} induced cycle of length at least 4; disoriented-chordal graphs are a subclass of this.}

\begin{figure}[h]
    \centering
    \includegraphics[width=\linewidth]{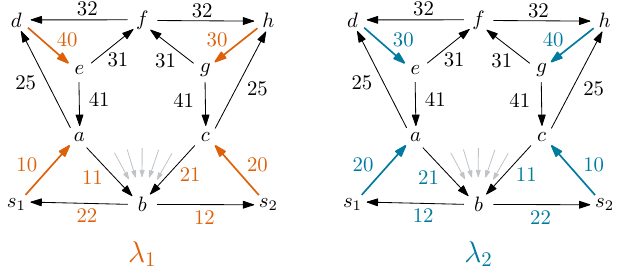}
    \caption{A disoriented-chordal graph with labelings that cannot be reconfigured into each other. Edges from $d, e, f, g, h$ towards $b$ are omitted for clarity, and their time labels can be arbitrary. Differing time labels and the four relevant edges mentioned in the proof below are colored.}
    \label{fig:2SR_chordal_no_instance}
\end{figure}

We disprove the hypothesis that disoriented-chordal graphs guarantee reconfiguration via the example in \Cref{fig:2SR_chordal_no_instance}. Note that $\lambda_2$ is a mirrored version of $\lambda_1$. This instance generalizes to any number of vertices by adding vertices with one incoming high-labeled edge from an original vertex. 

First verify that the disoriented graph is chordal: If we remove $b$, the graph is chordal. By adding $b$, we create new cycles, however $b$ is adjacent to all other vertices, so any cycle of length at least $4$ involving $b$ has a chord.

Verify that $\lambda_1$ and $\lambda_2$ both satisfy \sourcereach{2}. We now argue why $\lambda_1$ cannot be reconfigured into $\lambda_2$ wrt \sourcereach{2}.
For any labeling $\lambda$ satisfying \sourcereach{2}, we first show that if $\lambda((s_1, a)) < \lambda((s_2, c))$, then we must have $\lambda((h, g)) < \lambda((d, e))$ (as is the case in $\lambda_1$): Assume that $\lambda((s_1, a)) < \lambda((s_2, c))$. Then $s_2$ cannot reach $a$ via $(s_1, a)$. Thus the only other path from $s_2$ to $a$ that does not use $(s_1, a)$ is $s_2, c, h, g, f, d, e, a$. To make this path temporal, we must in particular have $\lambda((h, g)) < \lambda((d, e))$.

Since the graph is symmetric, the same argument shows that if $\lambda((s_2, c)) < \lambda((s_1, a))$, then we must have $\lambda((d, e)) < \lambda((h, g))$ to satisfy \sourcereach{2} (as is the case in $\lambda_2$).

This implies that there is no valid labeling with $\lambda((s_1, a)) < \lambda((s_2, c))$ and $\lambda((d, e)) < \lambda((h, g))$, and no valid labeling with $\lambda((s_2, c)) < \lambda((s_1, a))$ and $\lambda((h, g)) < \lambda((d, e))$. To reconfigure $\lambda_1$ into $\lambda_2$, one intermediate reconfiguration step must have such a labeling since a single reconfiguration step could only flip the order of one of the pairs. Hence reconfiguration is not possible.

\subsection{Almost-Tournament Graphs}

Disoriented-chordal graphs do not appear to enforce the right chords to prevent the conflicts that block reconfiguration. We further constrain the structure of cycles by forcing \emph{every} chord to exist, that is, there is exactly one edge between any two vertices (called a \emph{tournament graph}).
With a tournament graph as the footprint, reconfiguration wrt \sourcereach{2} is always possible simply following from the existence of an edge between $s_1$ and $s_2$: This edge allows one source to delegate to the other at time step $1$, thus only the other source's reachability needs to be maintained.

\begin{lemma}
    Given a graph $G=(V\cup \{s_1, s_2\}, E)$ where $s_1$ and $s_2$ are adjacent, we have $\lambda_1\reconf{\sourcereach{2}}\lambda_2$ for any labelings $\lambda_1, \lambda_2$ satisfying \sourcereach{2}.
\end{lemma}
\begin{proof}
    Let wlog $(s_1, s_2)\in E$. In both labelings, one can shift up all labels by $1$ and then change the label of $(s_1, s_2)$ to $0$, which does not break \sourcereach{2}. As long as $s_2$ reaches all vertices from time step $1$,  $s_1$ shares all these reachabilities by first taking the edge to $s_2$. If we now ignore $s_1$ and the edge $(s_1, s_2)$, we get a reconfiguration instance for $\sourcereach{1}$ where $s_2$ is the source, which is reconfigurable by \Cref{thm:1source_reach}.
\end{proof}

Therefore, we instead look at \emph{almost-tournament graphs}, where there is exactly one directed edge between any pair of vertices \emph{except between $s_1$ and $s_2$}. To prove that reconfiguration is still always possible, we will first separate the vertices in $V$ based on the edges between them and the sources:

\newcommand{\Vrr}{\ensuremath{V^{\rightarrow\rightarrow}}\xspace}
\newcommand{\Vlr}{\ensuremath{V^{\leftarrow\rightarrow}}\xspace}
\newcommand{\Vrl}{\ensuremath{V^{\rightarrow\leftarrow}}\xspace}
\newcommand{\Vll}{\ensuremath{V^{\leftarrow\leftarrow}}\xspace}

\begin{itemize}
    \item For $v\in \Vrr$, we have the edges $(s_1, v)$ and $(v, s_2)$.
    \item For $v\in \Vlr$, we have the edges $(v, s_1)$ and $(v, s_2)$.
    \item For $v\in \Vrl$, we have the edges $(s_1, v)$ and $(s_2, v)$.
    \item For $v\in \Vll$, we have the edges $(v, s_1)$ and $(s_2, v)$.
\end{itemize}

We initially shift up all labels in $\lambda_1$ and $\lambda_2$ by $|E|$, meaning there are no labels lower than $|E|$. This ensures that in later argumentation we are mainly \emph{reducing} time labels, so we only need to argue that they remain traversable from the sources.
To simplify arguments later, we will first reduce away all the vertices in \Vrl.

\begin{lemma}\label{lem:tournament_no_rightleft}
    Given an almost-tournament graph $G=(V\cup \{s_1, s_2\}, E)$, we either have $\lambda_1\reconf{\sourcereach{2}}\lambda_2$ for any labelings $\lambda_1, \lambda_2$ satisfying \sourcereach{2} or we can reduce to an equivalent instance with empty $\Vrl$.%
\end{lemma}
\begin{proof}
    
    We first reconfigure both $\lambda_1$ and $\lambda_2$ such that all edges between sources and vertices in \Vrl have label $1$: For any vertex $v\in \Vrl$, pick among $(s_1, v)$ and $(s_2, v)$ the edge with the larger label (resolve ties arbitrarily) and reduce its label to 1. Doing so maintains \sourcereach{2} as the incident source can still traverse the edge and does not arrive later at $v$, and the other source could previously reach any vertex without using that edge (its own edge to $v$ is at least as early). Afterwards, reduce the other edge to $1$ via the same argument.

    Now we reconfigure the other edges in $\lambda_1$ and $\lambda_2$ to be the same. We do so by running a breadth-first-search (BFS) on $G$ starting simultaneously from all vertices in \Vrl, traversing all reachable edges with label at least $|E|$. Each traversed edge is labeled with the BFS depth + 1 (so starting with 2). Whenever we reduce a label in this way, we do not break reachability as both sources can still traverse the edge by following the increasingly labeled BFS path, and they arrive earlier at the endpoint than they did before.

    If we traverse all edges this way, $\lambda_1$ and $\lambda_2$ will be identical afterwards. Otherwise, all the traversed edges are labeled identically in $\lambda_1$ and $\lambda_2$, and none of them can reach the remaining graph. Thus, if the remaining graph can be reconfigured, then the overall graph can be reconfigured.
\end{proof}

Assuming that $\Vrl$ is empty, we can show that reconfiguration is always possible.

\begin{lemma}\label{lem:tournament_general}
    Given an almost-tournament graph $G=(V\cup \{s_1, s_2\}, E)$ where $\Vrl$ is empty, we have $\lambda_1\reconf{\sourcereach{2}}\lambda_2$ for any labelings $\lambda_1, \lambda_2$ satisfying \sourcereach{2}.
\end{lemma}
\begin{proof}
    First note that neither $\Vrr$ nor $\Vll$ can be empty, or else one of the sources would have no outgoing edges.

    In $\lambda_1$ and $\lambda_2$, look at all edges from sources to vertices in $\Vll \cup \Vrr$ and pick the one with the smallest label (resolve ties arbitrarily). Let this be $(s_a, a)$ in $\lambda_1$ and $(s_b, b)$ in $\lambda_2$. We distinguish two cases:

    \bigparagraph{$s_a\neq s_b$:} So the earliest such edge goes out from different sources in both graphs. We assume wlog that the edge between $a$ and $b$ is $(b, a)$. %
    We now reconfigure $\lambda_1$ into a labeling $\lambda_1'$ as follows (see also \Cref{fig:tournament_diff_sources}):
    
    \begin{figure}[h]
        \centering
        \includegraphics[width=0.95\linewidth]{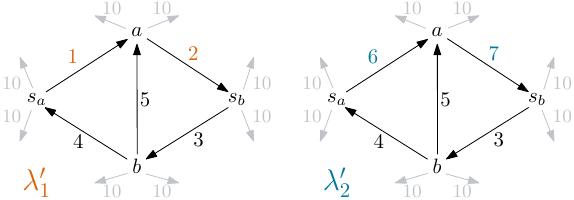}
        \caption{$\lambda_1'$ and $\lambda_2'$ if $s_a\neq s_b$. Differing labels are colored.}
        \label{fig:tournament_diff_sources}
    \end{figure}
    
    \begin{enumerate}
        \item Set the label of $(s_a, a)$ to $1$. It stays traversable from $s_a$, and $s_b$ was not able to traverse $(s_a, a)$ before as it had a smaller label than any edge outgoing from $s_b$.
        \item Set the label of $(a, s_b)$ to $2$. It remains traversable from $s_a$ via $(s_a, a)$ and is not needed by $s_b$. %
        \item Consecutively set the labels of $(s_b, b)$ and $(b, s_a)$ to $3$ and $4$ respectively. Both remain traversable from $s_b$ and thus also from $s_a$ via the edges labeled in the previous steps.
    \end{enumerate}

    Likewise, reconfigure $\lambda_2$ into $\lambda_2'$, but swapping the use of $a$ and $b$ and increasing assigned labels by $2$. Then set the label of $(b, a)$ to $5$ in both.
    Now all vertices in the subgraph $\{s_a, s_b, a, b\}$ can be reached by either source before time step 10. All edges outside the component can therefore be relabeled incrementally along a BFS outgoing from the component, starting with label 10.

    $\lambda_1'$ and $\lambda_2'$ now only differ in the edges $(s_a, a)$ and $(a, s_b)$. In $\lambda_2'$, first relabel $(s_a, a)$ to $1$; $s_a$ can naturally traverse it, and $s_b$ could already reach $a$ without using $(s_a, a)$ due to $(b, a)$. Then relabel $(a, s_b)$ to $2$; it remains traversable from $s_a$ and is not needed by $s_b$. This reconfigures $\lambda_2'$ into $\lambda_1'$, so overall we have $\lambda_1\reconf{\sourcereach{2}}\lambda_1'\reconf{\sourcereach{2}}\lambda_2'\reconf{\sourcereach{2}}\lambda_2$.
    
    \bigparagraph{$s_a = s_b$:} So the earliest such edge goes out from the same source in both graphs. Let this wlog be $s_1$. Now choose any $c\in \Vll$ (we showed this set was not empty). We distinguish two cases:
    \begin{itemize}
        \begin{figure}[h]
            \centering
            \includegraphics[width=0.95\linewidth]{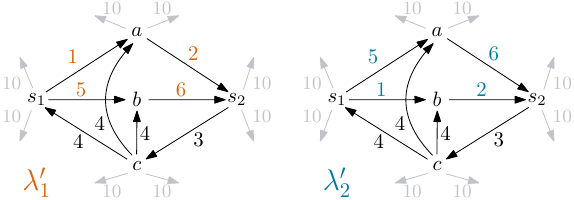}
            \caption{$\lambda_1'$ and $\lambda_2'$ if $s_a = s_b=s_1$ and the chosen $c$ has edges to $a$ and $b$. Differing labels are colored.}
            \label{fig:tournament_same_source1}
        \end{figure}
    
        \item If both $(c, a)$ and $(c, b)$ are edges, we reconfigure $\lambda_1$ and $\lambda_2$ into $\lambda_1'$ and $\lambda_2'$ as shown in \Cref{fig:tournament_same_source1}. We do so by applying the labels of $\lambda_1'$ (resp. $\lambda_2'$) from smallest to largest, and again labeling all edges not shown via a BFS. Each reconfiguration step is valid by the same argumentation as in the previous case.

        Now, in $\lambda_2'$ we relabel $(s_1, a)$ with $1$, then $(a, s_2)$ with $2$, then $(s_1, b)$ with $5$, and then $(b, s_2)$ with $6$ to make it equal to $\lambda_1'$. We get the reconfiguration sequence $\lambda_1\reconf{\sourcereach{2}}\lambda_1'\reconf{\sourcereach{2}}\lambda_2'\reconf{\sourcereach{2}}\lambda_2$.

        \begin{figure}[h]
            \centering
            \includegraphics[width=0.95\linewidth]{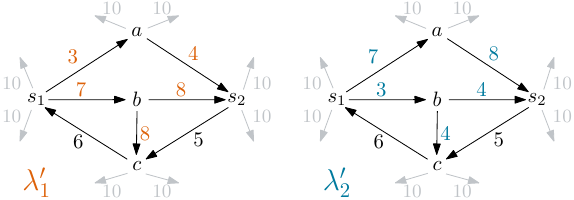}
            \caption{$\lambda_1'$ and $\lambda_2'$ if $s_a = s_b=s_1$ and $b$ has an edge to the chosen $c$. Differing labels are colored.}
            \label{fig:tournament_same_source2}
        \end{figure}
        \item Otherwise, at least one of $(a, c)$ or $(b, c)$ must be an edge, wlog let $(b, c)$ be an edge. We reconfigure $\lambda_1$ and $\lambda_2$ into $\lambda_1'$ and $\lambda_2'$ as shown in \Cref{fig:tournament_same_source2}, again applying labels from smallest to largest.

        Here however, we will reconfigure both $\lambda_1'$ and $\lambda_2'$ into a labeling $\lambda_c$ where $s_2$ delegates to $s_1$: In both labelings, relabel $(s_2, c)$ to 1 (the edge $(b, c)$ retains reachability from $s_1$ to $c$), then relabel $(c, s_1)$ to 2. Now relabel the remaining edges via a BFS from $s_1$, starting from label 3. We then have  $\lambda_1\reconf{\sourcereach{2}}\lambda_1'\reconf{\sourcereach{2}}\lambda_c\reconf{\sourcereach{2}}\lambda_2'\reconf{\sourcereach{2}}\lambda_2$.
    \end{itemize}

    Thus every case yields a reconfiguration $\lambda_1\reconf{\sourcereach{2}}\lambda_2$.
\end{proof}

\Cref{lem:tournament_no_rightleft,lem:tournament_general} together prove reconfigurability for almost-tournament graphs. Further note that our arguments are constructive and all described reconfiguration sequences (shifting up labels, labeling by BFS, resolving remaining differences) can be found in linear time.

\begin{theorem}
\label{thm:tournament_graphs}
    Given an almost-tournament graph $G=(V\cup \{s_1, s_2\}, E)$, we have $\lambda_1\reconf{\sourcereach{2}}\lambda_2$ for any labelings $\lambda_1, \lambda_2$ satisfying \sourcereach{2}, and a reconfiguration sequence can be computed in linear time.
\end{theorem}

\section{Conclusion}

Our results leave several open questions. The immediate ones are regarding this particular model. 
While we have identified graph classes where reconfiguration is always possible based on the graph's footprint, an interesting open question is whether there exists a graph property that exactly characterizes when every instance on a given footprint is reconfigurable.
Another natural direction is to identify properties of the initial and target labelings that guarantee the existence of a reconfiguration sequence.

More broadly, our work suggests a general framework for temporal graph reconfiguration under feasibility constraints. Besides temporal reachability, natural directions include preserving temporal strong connectivity, maintaining a bounded temporal diameter, or ensuring multiple disjoint temporal paths throughout the reconfiguration. Alternatively, one may study optimization objectives that deliberately restrict connectivity, such as minimizing temporal reachability, motivated by applications including livestock transportation networks, where movement schedules are updated while limiting the spread of infectious diseases.

\section*{Acknowledgements}

Georg Tennigkeit was supported by the HPI Research School on Foundations of AI (FAI).

\bibliography{references}

@inproceedings{ito2023reconfiguration,
  title={Reconfiguration of time-respecting arborescences},
  author={Ito, Takehiro and Iwamasa, Yuni and Kamiyama, Naoyuki and Kobayashi, Yasuaki and Kobayashi, Yusuke and Maezawa, Shun-ichi and Suzuki, Akira},
  booktitle={Algorithms and Data Structures Symposium},
  pages={521--532},
  year={2023},
  organization={Springer}
}

@inproceedings{dondi2024complexity,
  title={On the Complexity of Temporal Arborescence Reconfiguration},
  author={Dondi, Riccardo and Lafond, Manuel},
  booktitle={3rd Symposium on Algorithmic Foundations of Dynamic Networks (SAND 2024)},
  year={2024},
  organization={Schloss Dagstuhl--Leibniz-Zentrum f{\"u}r Informatik}
}

@article{Bonsma_2017,
    title={Rerouting shortest paths in planar graphs},
    volume={231},
    ISSN={0166218X},
    DOI={10.1016/j.dam.2016.05.024},
    journal={Discrete Applied Mathematics},
    author={Bonsma, Paul},
    year={2017},
    month=nov,
    pages={95–112},
    language={en} }

@article{DBLP:journals/tcs/KaminskiMM11,
  author       = {Marcin Kaminski and
                  Paul Medvedev and
                  Martin Milanic},
  title        = {Shortest paths between shortest paths},
  journal      = {Theor. Comput. Sci.},
  volume       = {412},
  number       = {39},
  pages        = {5205--5210},
  year         = {2011},
  url          = {https://doi.org/10.1016/j.tcs.2011.05.021},
  doi          = {10.1016/J.TCS.2011.05.021},
  bibsource    = {dblp computer science bibliography, https://dblp.org}
}

@article{DBLP:journals/talg/ItoIK0MNOO25,
  author       = {Takehiro Ito and
                  Yuni Iwamasa and
                  Naonori Kakimura and
                  Yusuke Kobayashi and
                  Shun{-}ichi Maezawa and
                  Yuta Nozaki and
                  Yoshio Okamoto and
                  Kenta Ozeki},
  title        = {Rerouting Planar Curves and Disjoint Paths},
  journal      = {{ACM} Trans. Algorithms},
  volume       = {21},
  number       = {2},
  pages        = {20:1--20:37},
  year         = {2025},
  url          = {https://doi.org/10.1145/3715694},
  doi          = {10.1145/3715694},
  bibsource    = {dblp computer science bibliography, https://dblp.org}
}

@article{Lettovsky2000,
  author    = {Libor Lettovsky and Ellis L. Johnson and George L. Nemhauser},
  title     = {Airline Crew Recovery},
  journal   = {Transportation Science},
  volume    = {34},
  number    = {4},
  pages     = {337--348},
  year      = {2000},
  doi       = {10.1287/trsc.34.4.337.12316}
}

@article{Clausen2010,
  author    = {Jens Clausen and Allan Larsen and Jesper Larsen and Olga Rezanova},
  title     = {Disruption Management in the Airline Industry---Concepts, Models and Methods},
  journal   = {Computers \& Operations Research},
  volume    = {37},
  number    = {5},
  pages     = {809--821},
  year      = {2010},
  doi       = {10.1016/j.cor.2009.03.027}
}

@article{Zhang2024,
  author  = {Tianyu Zhang and Gang Wang and Chuanyu Xue and
             Jiachen Wang and Mark Nixon and Song Han},
  title   = {Time-Sensitive Networking (TSN) for Industrial Automation:
             Current Advances and Future Directions},
  journal = {ACM Computing Surveys},
  volume  = {57},
  number  = {2},
  pages   = {30:1--30:38},
  year    = {2024},
  doi     = {10.1145/3695248}
}

@article{Gaertner2023,
  author  = {Christoph G{\"a}rtner and Amr Rizk and Boris Koldehofe and
             Ren{\'e} Guillaume and Ralf Kundel and Ralf Steinmetz},
  title   = {Fast Incremental Reconfiguration of Dynamic Time-Sensitive Networks at Runtime},
  journal = {Computer Networks},
  volume  = {224},
  pages   = {109606},
  year    = {2023},
  doi     = {10.1016/j.comnet.2023.109606}
}

@inproceedings{Gaertner2022,
  author    = {Christoph G{\"a}rtner and Amr Rizk and Boris Koldehofe and
               Ren{\'e} Guillaume and Ralf Kundel and Ralf Steinmetz},
  title     = {On the Incremental Reconfiguration of Time-sensitive Networks at Runtime},
  booktitle = {Proceedings of the IFIP Networking Conference},
  year      = {2022},
  publisher = {IFIP},
  doi        = {10.23919/IFIPNetworking55013.2022.9829815}
}

@inproceedings{DBLP:conf/aaai/GajjarJ0L22,
  author       = {Kshitij Gajjar and
                  Agastya Vibhuti Jha and
                  Manish Kumar and
                  Abhiruk Lahiri},
  title        = {Reconfiguring Shortest Paths in Graphs},
  booktitle    = {Thirty-Sixth {AAAI} Conference on Artificial Intelligence, {AAAI}
                  2022, Thirty-Fourth Conference on Innovative Applications of Artificial
                  Intelligence, {IAAI} 2022, The Twelveth Symposium on Educational Advances
                  in Artificial Intelligence, {EAAI} 2022 Virtual Event, February 22
                  - March 1, 2022},
  pages        = {9758--9766},
  publisher    = {{AAAI} Press},
  year         = {2022},
  url          = {https://doi.org/10.1609/aaai.v36i9.21211},
  doi          = {10.1609/AAAI.V36I9.21211},
  bibsource    = {dblp computer science bibliography, https://dblp.org}
}

@inproceedings{DBLP:conf/isaac/HanakaIKOS24,
  author       = {Tesshu Hanaka and
                  Yuni Iwamasa and
                  Yasuaki Kobayashi and
                  Yuto Okada and
                  Rin Saito},
  editor       = {Juli{\'{a}}n Mestre and
                  Anthony Wirth},
  title        = {Basis Sequence Reconfiguration in the Union of Matroids},
  booktitle    = {35th International Symposium on Algorithms and Computation, {ISAAC}
                  2024, Sydney, Australia, December 8-11, 2024},
  series       = {LIPIcs},
  volume       = {322},
  pages        = {38:1--38:16},
  publisher    = {Schloss Dagstuhl - Leibniz-Zentrum f{\"{u}}r Informatik},
  year         = {2024},
  url          = {https://doi.org/10.4230/LIPIcs.ISAAC.2024.38},
  doi          = {10.4230/LIPICS.ISAAC.2024.38},
  bibsource    = {dblp computer science bibliography, https://dblp.org}
}

@inproceedings{DBLP:conf/algosensors/DavotEL25,
  author       = {Tom Davot and
                  Jessica A. Enright and
                  Laura Larios{-}Jones},
  editor       = {Othon Michail and
                  Giuseppe Prencipe},
  title        = {Parameterised Algorithms for Temporally Satisfying Reconfiguration
                  Problems},
  booktitle    = {Algorithmics of Wireless Networks - 21st International Symposium,
                  {ALGOWIN} 2025, Warsaw, Poland, September 18-19, 2025, Proceedings},
  series       = {Lecture Notes in Computer Science},
  volume       = {16078},
  pages        = {89--103},
  publisher    = {Springer},
  year         = {2025},
  url          = {https://doi.org/10.1007/978-3-032-09120-8\_7},
  doi          = {10.1007/978-3-032-09120-8\_7},
  bibsource    = {dblp computer science bibliography, https://dblp.org}
}

@inproceedings{DBLP:conf/ijcai/DeligkasES23,
  author       = {Argyrios Deligkas and
                  Eduard Eiben and
                  George Skretas},
  title        = {Minimizing Reachability Times on Temporal Graphs via Shifting Labels},
  booktitle    = {Proceedings of the Thirty-Second International Joint Conference on
                  Artificial Intelligence, {IJCAI} 2023, 19th-25th August 2023, Macao,
                  SAR, China},
  pages        = {5333--5340},
  publisher    = {ijcai.org},
  year         = {2023},
  url          = {https://doi.org/10.24963/ijcai.2023/592},
  doi          = {10.24963/IJCAI.2023/592},
  bibsource    = {dblp computer science bibliography, https://dblp.org}
}

@article{DBLP:journals/tcs/ItoIKNOW23,
  author       = {Takehiro Ito and
                  Yuni Iwamasa and
                  Yasuaki Kobayashi and
                  Yu Nakahata and
                  Yota Otachi and
                  Kunihiro Wasa},
  title        = {Reconfiguring (non-spanning) arborescences},
  journal      = {Theor. Comput. Sci.},
  volume       = {943},
  pages        = {131--141},
  year         = {2023},
  url          = {https://doi.org/10.1016/j.tcs.2022.12.007},
  doi          = {10.1016/J.TCS.2022.12.007},
  bibsource    = {dblp computer science bibliography, https://dblp.org}
}

@article{DBLP:journals/tcs/EnrightLMP26,
  author       = {Jessica A. Enright and
                  Laura Larios{-}Jones and
                  Kitty Meeks and
                  William Pettersson},
  title        = {Reachability in temporal graphs under perturbation},
  journal      = {Theor. Comput. Sci.},
  volume       = {1083},
  pages        = {116138},
  year         = {2026},
  url          = {https://doi.org/10.1016/j.tcs.2026.116138},
  doi          = {10.1016/J.TCS.2026.116138},
  bibsource    = {dblp computer science bibliography, https://dblp.org}
}

@article{DBLP:journals/jcss/MolterRZ24,
  author       = {Hendrik Molter and
                  Malte Renken and
                  Philipp Zschoche},
  title        = {Temporal reachability minimization: Delaying vs. deleting},
  journal      = {J. Comput. Syst. Sci.},
  volume       = {144},
  pages        = {103549},
  year         = {2024},
  url          = {https://doi.org/10.1016/j.jcss.2024.103549},
  doi          = {10.1016/J.JCSS.2024.103549},
  bibsource    = {dblp computer science bibliography, https://dblp.org}
}

@article{DBLP:journals/iandc/DeligkasP22,
  author       = {Argyrios Deligkas and
                  Igor Potapov},
  title        = {Optimizing reachability sets in temporal graphs by delaying},
  journal      = {Inf. Comput.},
  volume       = {285},
  number       = {Part},
  pages        = {104890},
  year         = {2022},
  url          = {https://doi.org/10.1016/j.ic.2022.104890},
  doi          = {10.1016/J.IC.2022.104890},
  bibsource    = {dblp computer science bibliography, https://dblp.org}
}

@article{DBLP:journals/networks/HalpernP74,
  author       = {Jonathan Halpern and
                  I. Priess},
  title        = {Shortest path with time constraints on movement and parking},
  journal      = {Networks},
  volume       = {4},
  number       = {3},
  pages        = {241--253},
  year         = {1974},
  url          = {https://doi.org/10.1002/net.3230040304},
  doi          = {10.1002/NET.3230040304},
  bibsource    = {dblp computer science bibliography, https://dblp.org}
}

@article{DBLP:journals/ijfcs/XuanFJ03,
  author       = {Binh{-}Minh Bui{-}Xuan and
                  Afonso Ferreira and
                  Aubin Jarry},
  title        = {Computing Shortest, Fastest, and Foremost Journeys in Dynamic Networks},
  journal      = {Int. J. Found. Comput. Sci.},
  volume       = {14},
  number       = {2},
  pages        = {267--285},
  year         = {2003},
  url          = {https://doi.org/10.1142/S0129054103001728},
  doi          = {10.1142/S0129054103001728},
  bibsource    = {dblp computer science bibliography, https://dblp.org}
}

@article{DBLP:journals/algorithms/Nishimura18,
  author       = {Naomi Nishimura},
  title        = {Introduction to Reconfiguration},
  journal      = {Algorithms},
  volume       = {11},
  number       = {4},
  pages        = {52},
  year         = {2018},
  url          = {https://doi.org/10.3390/a11040052},
  doi          = {10.3390/A11040052},
  bibsource    = {dblp computer science bibliography, https://dblp.org}
}

@article{DBLP:journals/tcs/AkitayaKKST22,
  author       = {Hugo A. Akitaya and
                  Matias Korman and
                  Oliver Korten and
                  Diane L. Souvaine and
                  Csaba D. T{\'{o}}th},
  title        = {Reconfiguration of connected graph partitions via recombination},
  journal      = {Theor. Comput. Sci.},
  volume       = {923},
  pages        = {13--26},
  year         = {2022},
  url          = {https://doi.org/10.1016/j.tcs.2022.04.049},
  doi          = {10.1016/J.TCS.2022.04.049},
  bibsource    = {dblp computer science bibliography, https://dblp.org}
}

@article{DBLP:journals/dam/ItoOO23,
  author       = {Takehiro Ito and
                  Hirotaka Ono and
                  Yota Otachi},
  title        = {Reconfiguration of cliques in a graph},
  journal      = {Discret. Appl. Math.},
  volume       = {333},
  pages        = {43--58},
  year         = {2023},
  url          = {https://doi.org/10.1016/j.dam.2023.01.026},
  doi          = {10.1016/J.DAM.2023.01.026},
  bibsource    = {dblp computer science bibliography, https://dblp.org}
}

@inproceedings{DBLP:conf/hotnets/ReitblattFRW11,
  author       = {Mark Reitblatt and
                  Nate Foster and
                  Jennifer Rexford and
                  David Walker},
  editor       = {Hari Balakrishnan and
                  Dina Katabi and
                  Aditya Akella and
                  Ion Stoica},
  title        = {Consistent updates for software-defined networks: change you can believe
                  in!},
  booktitle    = {Tenth {ACM} Workshop on Hot Topics in Networks (HotNets-X), {HOTNETS}
                  '11, Cambridge, MA, {USA} - November 14 - 15, 2011},
  pages        = {7},
  publisher    = {{ACM}},
  year         = {2011},
  url          = {https://doi.org/10.1145/2070562.2070569},
  doi          = {10.1145/2070562.2070569},
  bibsource    = {dblp computer science bibliography, https://dblp.org}
}

@article{DBLP:journals/jcss/KempeKK02,
  author       = {David Kempe and
                  Jon M. Kleinberg and
                  Amit Kumar},
  title        = {Connectivity and Inference Problems for Temporal Networks},
  journal      = {J. Comput. Syst. Sci.},
  volume       = {64},
  number       = {4},
  pages        = {820--842},
  year         = {2002},
  url          = {https://doi.org/10.1006/jcss.2002.1829},
  doi          = {10.1006/JCSS.2002.1829},
  bibsource    = {dblp computer science bibliography, https://dblp.org}
}

@article{DBLP:journals/algorithmica/MertziosMS19,
  author       = {George B. Mertzios and
                  Othon Michail and
                  Paul G. Spirakis},
  title        = {Temporal Network Optimization Subject to Connectivity Constraints},
  journal      = {Algorithmica},
  volume       = {81},
  number       = {4},
  pages        = {1416--1449},
  year         = {2019},
  url          = {https://doi.org/10.1007/s00453-018-0478-6},
  doi          = {10.1007/S00453-018-0478-6},
  bibsource    = {dblp computer science bibliography, https://dblp.org}
}

\end{document}